\documentclass[sigconf]{acmart}
\usepackage{subfigure}
\usepackage{multirow}
\usepackage{color, verbatim,caption}
\usepackage{hyperref}
\usepackage{enumitem}
\usepackage{subfigure}
\usepackage{float}
\usepackage{placeins}

\usepackage{caption} 
\usepackage{algorithm}
\usepackage{algorithmicx}
\usepackage{algpseudocode}
\usepackage{amsmath}
\usepackage{xcolor}

\acmDOI{10.475/123_4}

\acmISBN{123-4567-24-567/08/06}

\begin{document}
\newcommand{\hiformer}{\textsc{Hiformer}~~}
\newcommand{\etal}{\textit{et al.}}

\definecolor{myRed}{rgb}{255,0,0}
\definecolor{myBlue}{rgb}{0, 0,255}
\definecolor{myPink}{rgb}{255, 0,255}


\title{Hiformer: Heterogeneous Feature Interactions Learning with Transformers for Recommender Systems}



\author{Huan Gui}
\affiliation{%
 \institution{Google DeepMind}
 \city{Mountain View}
 \state{California}
 \country{USA}}
\email{hgui@google.com}
\author{Ruoxi Wang}
\affiliation{%
 \institution{Google DeepMind}
 \city{Mountain View}
 \state{California}
 \country{USA}}
\email{ruoxi@google.com}
\author{Ke Yin}
\affiliation{%
 \institution{Google Inc}
 \city{Mountain View}
 \state{California}
 \country{USA}}
\email{keyin@google.com}
\author{Long Jin}
\affiliation{%
 \institution{Google Inc}
 \city{Mountain View}
 \state{California}
 \country{USA}}
\email{longjin@google.com}
\author{Maciej Kula}
\affiliation{%
 \institution{Google DeepMind}
 \city{Mountain View}
 \state{California}
 \country{USA}}
\email{maciejkula@google.com}
\author{Taibai Xu}
\affiliation{%
 \institution{Google Inc}
 \city{Mountain View}
 \state{California}
 \country{USA}}
\email{taibaixu@google.com}
\author{Lichan Hong}
\affiliation{%
 \institution{Google DeepMind}
 \city{Mountain View}
 \state{California}
 \country{USA}}
\email{lichan@google.com}
\author{Ed H. Chi}
\affiliation{%
 \institution{Google DeepMind}
 \city{Mountain View}
 \state{California}
 \country{USA}}
\email{edchi@google.com}

\begin{abstract} 
Learning feature interaction is the critical backbone to building recommender systems.
In web-scale applications, learning feature interaction is extremely challenging due to the sparse and large input feature space; meanwhile, manually crafting effective feature interactions is infeasible because of the exponential solution space. 
We propose to leverage a Transformer-based architecture with attention layers to automatically capture feature interactions.
Transformer architectures have  witnessed great success  in many domains, such as natural language processing and computer vision.
However, there has not been much adoption of Transformer architecture for feature interaction modeling in industry.
We aim at closing the gap. 
We identify two key challenges for applying the vanilla Transformer architecture to web-scale recommender systems:
(1) Transformer architecture fails to capture the heterogeneous feature interactions in the self-attention layer;
(2) The serving latency of Transformer architecture might be too high to be deployed in web-scale recommender systems.
We first propose a heterogeneous self-attention layer, which is a simple yet effective modification to the self-attention layer in Transformer, to take into account the heterogeneity of feature interactions.
We then introduce \textsc{Hiformer} (\textbf{H}eterogeneous \textbf{I}nteraction Trans\textbf{former}) to further improve the model expressiveness. With  low-rank approximation and model pruning, \hiformer  enjoys fast inference  for online deployment.
Extensive offline experiment results corroborates the effectiveness and efficiency  of the \textsc{Hiformer} model.
We have successfully deployed the \textsc{Hiformer} model to a real world large scale App ranking model at Google Play, with significant improvement in key engagement metrics (up to +2.66\%). 

\end{abstract}

\terms{Algorithms, Experimentation}


\maketitle

\section{INTRODUCTION}
\label{sec::intro}

The internet is inundated with a plethora of information, making it challenging for users to effectively navigate and locate relevant content.
Recommender systems filter information and present the most relevant content to individual users~\cite{resnick1997recommender,zhang2019deep}.
It is common to formulate recommender systems  as a supervised machine learning problem, with the goal of increasing user's positive engagements with the recommendations, such as clicks~\cite{wang2017deep, wang2021dcn, song2019autoint}, watches~\cite{zhao2019recommending}, or purchases~\cite{schafer1999recommender}.
Therefore, the deployment of recommender systems with high prediction accuracy is of paramount importance as it  could have a direct influence on business' financial performance, such as sales and revenue~\cite{zhang2019deep, cheng2016wide, covington2016deep, davidson2010youtube}.


Feature interactions are where multiple features have complex collaborative effects on the prediction of an outcome~\cite{tsang2017detecting, wang2021dcn}.
It is one of the crucial components of recommender system.
For example, it's observed that users frequently download food delivery apps during meal times~\cite{guo2017deepfm}, indicating the feature interactions of app ids (\textit{e.g.} food delivery apps) and temporal context (\textit{e.g.} meal times) can provide vital signal for predicting user behavior and making personalized recommendations.
However, modeling feature interactions in web-scale recommender systems presents three significant challenges.
First of all, the right feature interactions are often domain-specific, manually crafting the feature interactions is time-consuming and requires domain knowledge. 
Secondly, for web-scale applications, due to the large number of raw features, the search of possible feature interactions has an exponential solution space. This makes it infeasible to manually extract all the interactions. 
Last but not least, the extracted feature interactions may not generalize well to other tasks. 
Due to its importance, feature interaction continues to attract increasing attention from both academic and industry~\cite{lu2015recommender, zhang2019deep, covington2016deep, davidson2010youtube, qu2018product}.

In recent years, deep neural networks (DNNs) have garnered significant attention in various research fields, due to their exceptional model performance as well as their ability in representation learning~\cite{lecun2015deep,zhang2019deep}. 
DNNs have been widely adopted for recommender system for sparse input features representation learning~\cite{okura2017embedding} and feature interaction learning~\cite{lian2018xdeepfm, qu2016product, song2019autoint, wang2021dcn, zhang2019deep, beutel2018latent, cheng2016wide, guo2017deepfm, naumov2019deep, wang2017deep}.
It has become a powerful tool to map the large and sparse input into a low-dimensional semantic space.
Some of the recent works~\cite{qu2016product, naumov2019deep, li2016difacto, wang2021dcn, guo2017deepfm} in feature interaction learning are  based on explicit feature interaction functions, such as Factorization Machine~\cite{rendle2010factorization, lian2018xdeepfm}, inner product~\cite{qu2016product}, kernel Factorization Machine~\cite{qu2018product},  and Cross Network~\cite{wang2017deep, wang2021dcn}. 
DNNs are also leveraged to learn the implicit feature interactions with multiple layers of artificial neurons with non-linear activation functions~\cite{cheng2016wide, naumov2019deep}.


Another line of work is based on the attention mechanism, in particular, the  Transformer-based architecture~\cite{vaswani2017attention} for feature interaction learning.
The Transformer architecture has become the state-of-the-art (SOTA) model for a variety of tasks, including  computer vision (CV)~\cite{han2022survey, khan2022transformers}  and   natural language processing (NLP)~\cite{kalyan2021ammus}.
AutoInt~\cite{song2019autoint} and InterHAt~\cite{li2020interpretable}  proposed to leverage the multi-head self-attention layer for feature interaction learning.
However, unlike the NLP field, the feature semantics in recommender systems are dynamic under different contexts. 

Suppose we are recommending food delivery apps to users and we have the features: \texttt{app\textunderscore{}id}, \texttt{hour\textunderscore{}of\textunderscore{}day}, and \texttt{user\textunderscore{}country}. In this case, \texttt{app\textunderscore{}id} could mean one thing for \texttt{hour\textunderscore{}of\textunderscore{}day} and another for \texttt{user\textunderscore{}country}.
To effectively interact different features, we would need semantic awareness and semantic space alignment. Transformer models unfortunately do not consider this. In the vanilla attention layer, the features are projected through the same projection matrices (\textit{i.e.,} $\mathbf{W_Q}$ and $\mathbf{W_k}$) that are shared across all features. This is a natural design for applications where the feature (text token) semantics are independent of the context; however for recommender systems where the feature semantic are often dependent of the context, this homogeneous design would lead to limited model expressiveness for feature interaction learning.

Besides the limitation of not being able to capture heterogeneous feature interactions, another limitation of Transformer models is on latency. In web-scale applications, there are frequently a vast number of users and requests. 
It is crucial that the models are capable of managing an extremely high number of queries per second, to deliver an optimal user experience.
However, the inference cost of Transformer architectures scales quadratically with the input length, which makes real-time inference using Transformer architectures prohibitive in web-scale recommender systems.

Despite the limitations, we deem it important to study and improve Transformer-based architecture for feature interaction. 
First of all, due to the success of the Transformer architecture in many domains, a significant number of advancements and modifications have been made to the architecture to improve its performance and capabilities.
Enabling Transformer-based feature interaction models for recommender system can serve as a bridge, to allow for the recent advancements in Transformer architectures to be applied in the recommendation domain.
Secondly, with its widespread adoption, the hardware design (e.g., chip design~\cite{mirhoseini2021graph}) might favor Transformer-like architectures. Hence, recommender systems with  Transformer-based architectures could also enjoy the  optimizations brought by hardware.
Thirdly, as Transformer models for feature interaction learning are based on the attention mechanism, they provide  good model explainability~\cite{song2019autoint}. 
Building recommender models with attention mechanism in web-scale applications might open up new research opportunities. 

In this paper, we consider the limitations of vanilla Transformer model architectures and propose a new model:  \hiformer~ (\textbf{H}eterog-eneous Feature \textbf{I}nteraction Trans\textbf{former}) for feature interaction learning in recommender systems. \hiformer~ is feature semantic aware and, more importantly, efficient for online deployment. 
To provide feature semantic awareness in feature interaction learning, we design a novel heterogeneous attention layer.
To comply with the stringent latency requirements of web-scale applications, we leverage low-rank approximation and model pruning techniques to improve the efficiency.


We use a large-scale App ranking model at Google Play as a case study.
We perform both offline and online experiments with \hiformer  to corroborate its effectiveness in feature interaction modeling and efficiency in serving. 
To the best of our knowledge, we show for the first time that a Transformer-based architecture (i.e.,  \hiformer)  can outperform SOTA recommendation models for feature interaction learning.
We have successfully deployed  \hiformer as the  production model.

To summarize, the contributions of the paper are as follows:
\begin{itemize}
    \item We propose the \hiformer model with a novel heterogeneous attention layer to capture the complex collaborative effects of features in feature interaction learning. Compared with existing vanilla Transformer based approaches, our model has  more model   expressiveness.
    \item We leverage  low-rank approximation and model pruning to reduce the serving latency of the \hiformer model, without compromising the model quality.
    \item We conduct extensive offline comparison with a web-scale dataset to demonstrate the importance of capturing heterogeneous feature interactions. We show that \textsc{hiformer}, as a Transformer-based architecture, can outperform SOTA recommendation models.
    \item We  perform online A/B testing to  measure the impact of different models on key  engagement metrics, and  the \textsc{Hiformer} model shows significant online gain compared with the baseline model with limited latency increase.
\end{itemize}





\section{PROBLEM DEFINITION}
\label{sec::definition}

The goal of recommender system is to improve users' positive engagement, such as clicks, conversion, etc.
Similar to previous studies~\cite{lian2018xdeepfm, lu2015recommender}, we formulate it as a supervised machine learning task for engagement predictions.

\begin{definition}[Recommender System]
Let $\textbf{x} \in\mathbb{R}^{d_x}$  denote the features of a  (user, item) pair.
There are categorical features $\textbf{x}^\mathcal{C}$ and dense features  $\textbf{x}^\mathcal{D}$.
Categorical features are represented with one-hot encoding.
The dense features can be considered as special embedding features with embedding dimension being 1.
The numbers of categorical and dense features  are $|\mathcal{C}|$ and $|\mathcal{D}|$ respectively.
The problem of recommender system is to predict if the user will engage with the item based on input features $\textbf{x}$.
\end{definition}

As aforementioned, feature interaction learning enables the model to capture more complex relationships among features, thereby provides virtual information for more accurate recommendations.
We formally define heterogeneous feature interaction learning.

\begin{definition}[Heterogeneous $z$-order Feature Interaction]
\label{def:heterogeneous-feature-interaction}
Given input with $|\mathcal{F}|$ features and the input as $\mathbf{x} = \{\mathbf{x}_{i}\}^{|\mathcal{F}|}_{i=1}$, a $z$-order feature interaction is to learn a unique \textit{non-additive} mapping function $\rho^{\mathcal{Z}}(\cdot)$ that map from the feature list $\mathcal{Z}$ with $z$ features (\textit{i.e.,} $[\mathbf{x}_{i_1}, \cdots \mathbf{x}_{i_z} ]$) to a new representation or a scalar to capture the complex relationships among the the feature list $\mathcal{Z}$.
When z = 2,  $\rho^{\mathcal{Z}}(\cdot)$ captures the second order feature interactions.
\end{definition}

For example, $g([\mathbf{x}_1, \mathbf{x}_2]) = w_1 \mathbf{x}_1 +  w_2 \mathbf{x}_2 $ is not a 2-order feature interaction, as  $w_1 \mathbf{x}_1 +  w_2 \mathbf{x}_2$ is additive. 
Meanwhile, $g([\mathbf{x}_1, \mathbf{x}_2]) = \mathbf{x_1}^T\mathbf{x_2} $ is a 2-order feature interaction; and $\rho^{1,2}(\cdot) = \rho^{2,1}(\cdot) = \mathbf{x}_1^T\mathbf{x}_2$, meaning the heterogeneous feature interaction is symmetric. 

Before diving into model details, we define our notations. 
Scalars are denoted by lower case letters $(a, b, . . .)$, vectors by bold lower case letters $(\textbf{a}, \textbf{b}, . . .)$,
matrices by bold upper case letters $(\textbf{A}, \textbf{B}, . . .)$.


\section{MODEL}
\label{sec::framework}

\begin{figure}[tp]
    \centering
    \includegraphics[width=0.5\textwidth]{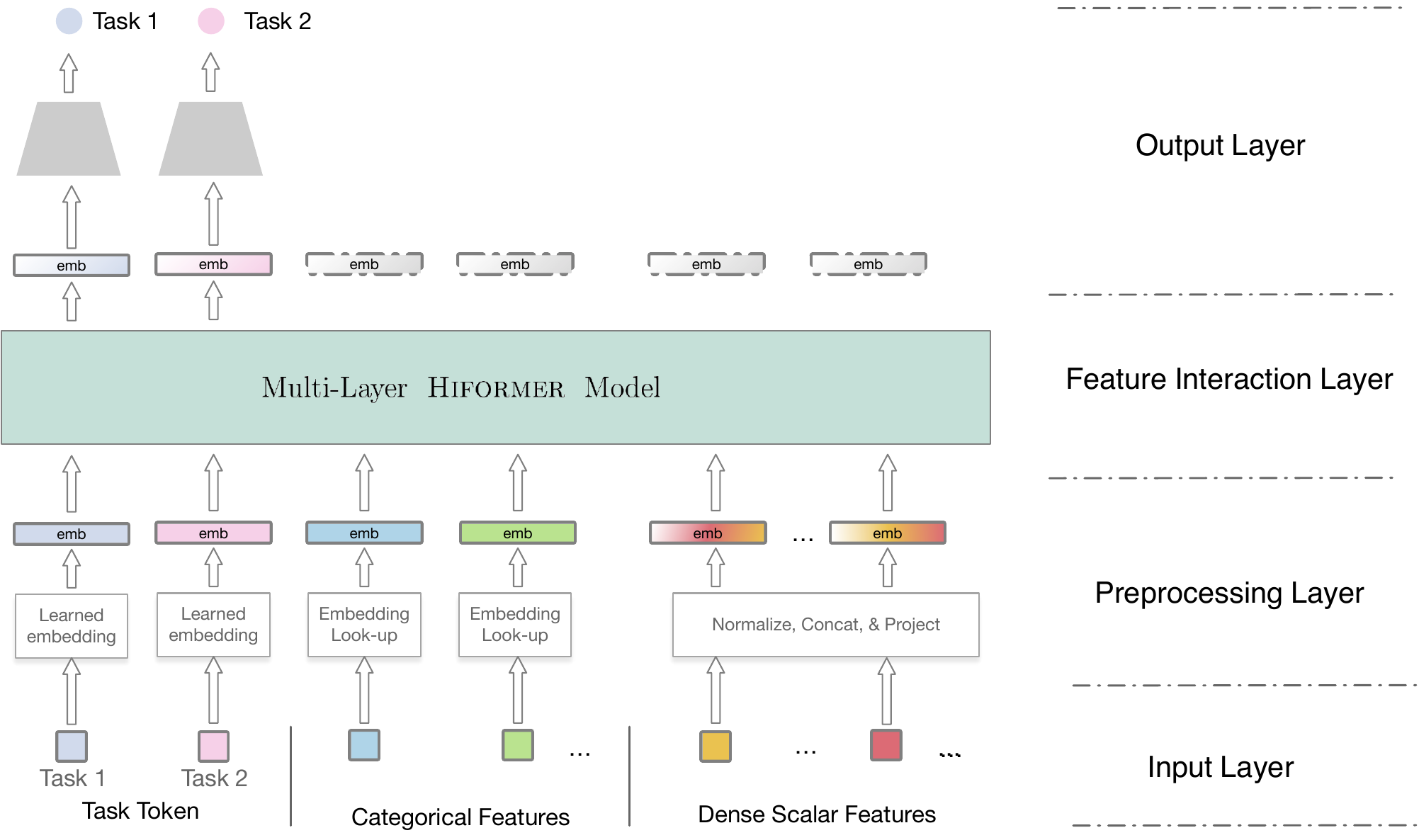}
    
    \vspace{-3mm}
    \caption{There are four components in the framework: Input Layer, Preprocessing Layer, Feature Interaction layer, and Output Layer. We leverage a novel \hiformer~~model  for heterogeneous feature interaction learning. }\label{visina8}%
    \label{fig:framework}
    \vspace{-4mm}
\end{figure}





We will first present the overall framework  in Section~\ref{sec::framework-overview}.
We then propose a heterogeneous attention layer in Section~\ref{sec::hetero-att}, based on which we further introduce the \hiformer model  in Section~\ref{sec::hiformer}. 

\subsection{Overview}
\label{sec::framework-overview}


\subsubsection{Input Layer.} The input layer includes the categorical features, and dense features. 
Additionally, we have the task embedding. 
The task embedding can be regarded as the CLS token~\cite{devlin2018bert}. 
The proposed framework can also be easily extended to multi-task~\cite{caruana1997multitask} with multiple task embeddings, and each task embedding represents information of the corresponding training objective.
The task embeddings in the input layers are model parameters and are learned through the end-to-end model training. 
We denote the number of task embeddings as $t$.

\subsubsection{Preprocessing Layer}
We have the preprocessing layer to transform the input layer into a list of feature embeddings, in addition to the task embeddings, as input to the feature interaction layers. 
We have one preprocessing layer per feature type.

\textbf{Categorical Features.} 
The input categorical features are often sparse and of high dimensions. 
Directly using the one-hot encoding for model training would easily lead to model overfitting. 
Therefore,  a common strategy is to project the categorical features into a low-dimension dense space. 
In particular, we learn a projection matrix $\mathbf{W}_i^{\mathcal{C}} \in \mathbb{R}^{V_i \times d}$ related to each categorical feature, and 
\begin{equation*}
    \mathbf{e}_i = \mathbf{x}_i^{\mathcal{C}} \mathbf{W}_i^{\mathcal{C}},
\end{equation*}
where $\mathbf{e}_i \in \mathbb{R}^d$,  $\mathbf{x}_i^{\mathcal{C}} \in \{0, 1\}^{V_i}$, $V_i$ is dimensions of the one-hot encoding for feature $\mathbf{x}_i^{\mathcal{C}} $, and $d$ is the model dimension.
We will then use the dense embedded vector $\mathbf{e}_i$ as the representation of the categorical feature $x_i^\mathcal{C}$.
$\{\textbf{W}_i^{\mathcal{C}}\}^{|\mathcal{C}|}$ are the model parameters and referred to as  embedding look-up tables.


\textbf{ Dense Scalar Features.} 
The dense scalar features are numerical features.
As the dense scalar features are potentially from very different distributions, we will need to transform the features to a similar distribution (\textit{i.e.,} uniform distribution) to ensure numerical stability~\cite{zhuang2020feature}.
Additionally, in web-scale applications, there are  a large number of dense scalar features, with much less compared with the categorical features.  
Therefore, we aggregate all the dense scalar features and project them into $n^\mathcal{D}$ embeddings~\cite{naumov2019deep}, where $n^\mathcal{D}$ is a hyper-parameter chosen by empirical results, $n^\mathcal{D} \ll |\mathcal{D}|$ to reduce the total number of features.
The projection function $f_\mathcal{D}(\cdot)$ could be a  multilayer perceptron (MLP) with nonlinear activations:
\begin{equation*}
    \mathbf{e}_i = \text{split}_i \bigg( f_\mathcal{D}\bigg (\text{concat} \big(\text{normalize}(\{x_i^{\mathcal{D}}\})\big) \bigg), \texttt{split\textunderscore{}size}=d \bigg),
\end{equation*}
where $\mathbf{e}_i \in \mathbb{R}^d$, $\text{concat}(\cdot)$ is the concatenation function, $\text{normalize}(\cdot)$ is the feature transformation function,  and 
$\text{split}_i(\cdot, \texttt{split\textunderscore{}size})$ is a function to split the input tensor into tensors with equal dimension of  $\texttt{split\textunderscore{}size}$ and takes the $i$-th tensor as output.
In Transformer-based architectures, the inference cost scales quadratically with the length of input embedding list.
By aggregating $|\mathcal{D}|$ dense scalar features into $n^\mathcal{D}$ embeddings, the length of feature embeddings  is reduced, and as a consequence, the inference cost is also reduced, albeit with a trade-off of adding $f_D(\cdot)$.


To summarize on preprocessing layers, we transform the $|\mathcal{C}|$ categorical features,  $|\mathcal{D}|$ dense scalar features, and the $t$ task embeddings into a list of embeddings with the model dimension $d$.
The number of output embeddings from preprocessing layers is
$L = |\mathcal{C}| + n^\mathcal{D} + T$. 
For ease of discussion, the task embeddings are also regarded as special features in the embedding list.
Therefore, feature interactions refers to  not only   interaction among features, but also between features and tasks.


\subsubsection{Feature Interaction Layer}
The feature interaction layer is used to learn the complex collaborative effects among features, which is crucial to capture users' preferences under different contexts.
The input to the feature interaction layer is the output embedding list of preprocessing layers.


\subsubsection{Output Layer and Training Objectives} We only use the encoded task embeddings from the feature interaction layer for task prediction.
We train a MLP tower to project the encoded task embedding to final predictions. 
The training tasks can be flexibly configured based on the recommendation application. 
For example, the training task can be a classification task if the labels are click~\cite{song2020towards, song2019autoint}, installation~\cite{yang2020mixed, joglekar2020neural}, or purchase~\cite{pan2022metacvr, li2021attentive}, and may also be a regression task, such as in the case of watch time~\cite{zhao2019recommending}. 
Without loss of generality, we consider classification task, with a value of 0 indicating negative engagement and  1 indicating positive engagement.
Thus, the training objective can be formulated as a binary classification problem with the Binary Cross Entropy (\textit{i.e.}, Log Loss):
\begin{equation*}
\ell = \frac{1}{N}\sum_{i}^N -y_i \log(p_i) - (1-y_i)\log(1-p_i),
    \label{eq:logloss}
\end{equation*}
where $y_i$ and $p_i$ are the ground truth label and predicted probability of engagements for example $i$ respectively, and $N$ is the total number of training examples.



\subsection{Heterogeneous Feature Interactions}
The Transformer architecture has become the de-facto standard for many tasks such as NLP, CV, etc., and multi-head self-attention layer has achieved remarkable performance in modeling complicated relations in sequences~\cite{vaswani2017attention, han2022survey,kalyan2021ammus,khan2022transformers}.
Nevertheless, the Transformer architecture has not been extensively applied to web-scale recommender systems.
It has been observed that Transformer-based architectures, such as AutoInt~\cite{song2019autoint}, may not demonstrate optimal performance when applied to web-scale recommender systems.

\begin{figure*}[tp]
    \centering
    \subfigure[Vanilla Transformer.]
    {
        \includegraphics[width=0.29\textwidth]{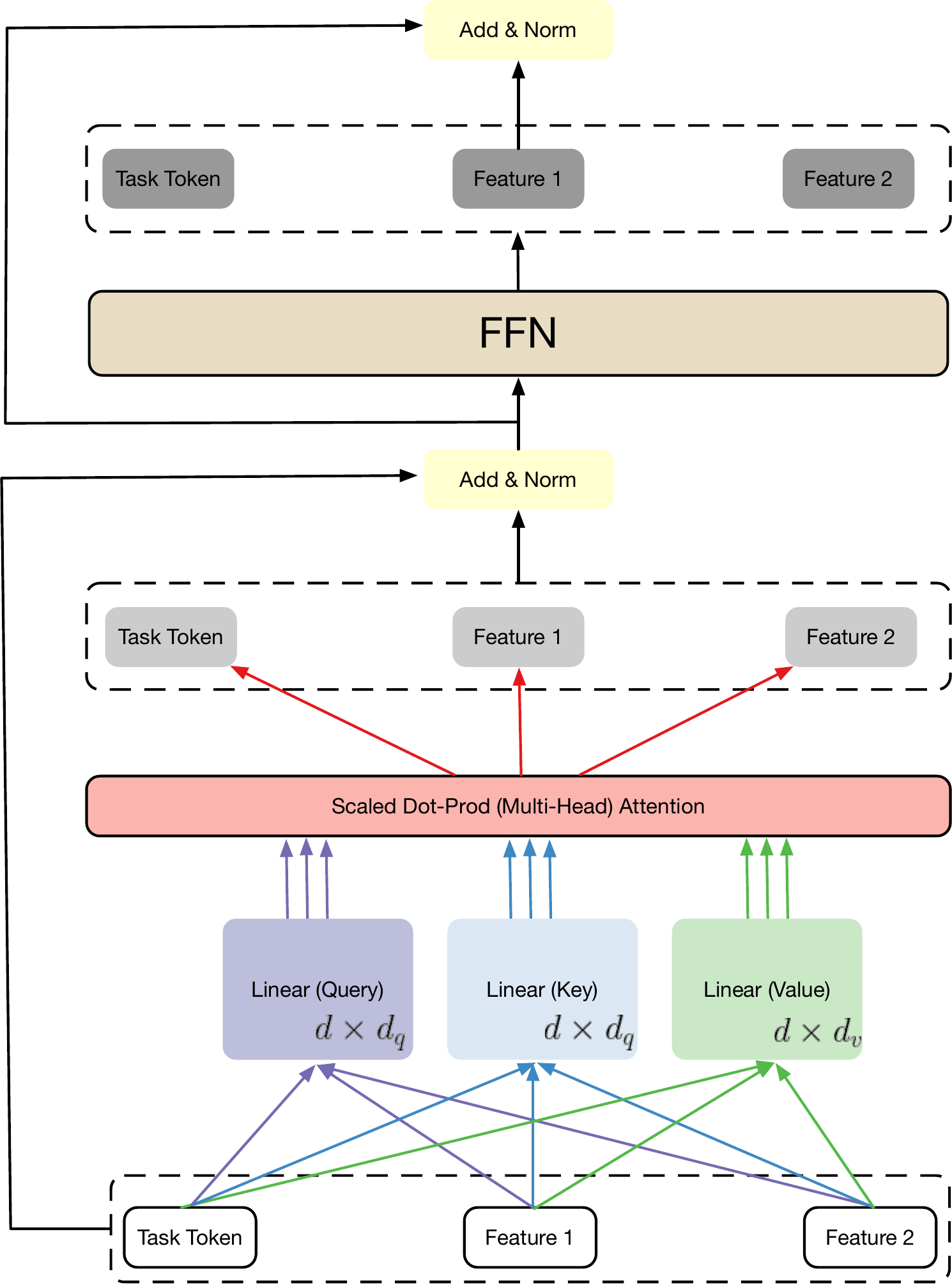}
        \label{fig:fiformer}
    }
      \hspace*{\fill} 
    \subfigure[Transformer with Heterogeneous Attention.]
    {
        \includegraphics[width=0.32\textwidth]{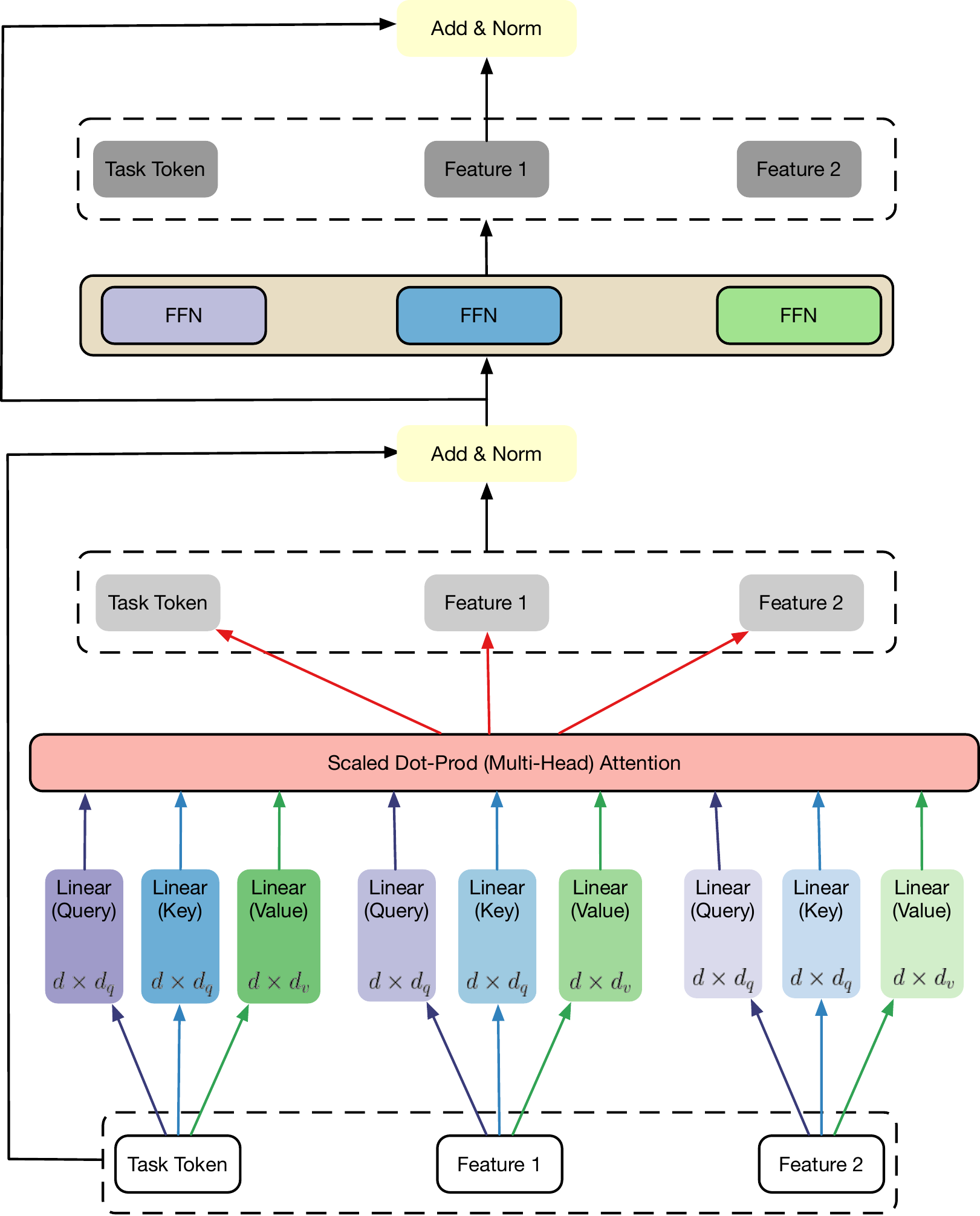}
        \label{fig:hiformer}
    }
      \hspace*{\fill} 
    \subfigure[\textsc{Hiformer}.]
    {
        \includegraphics[width=0.29\textwidth]{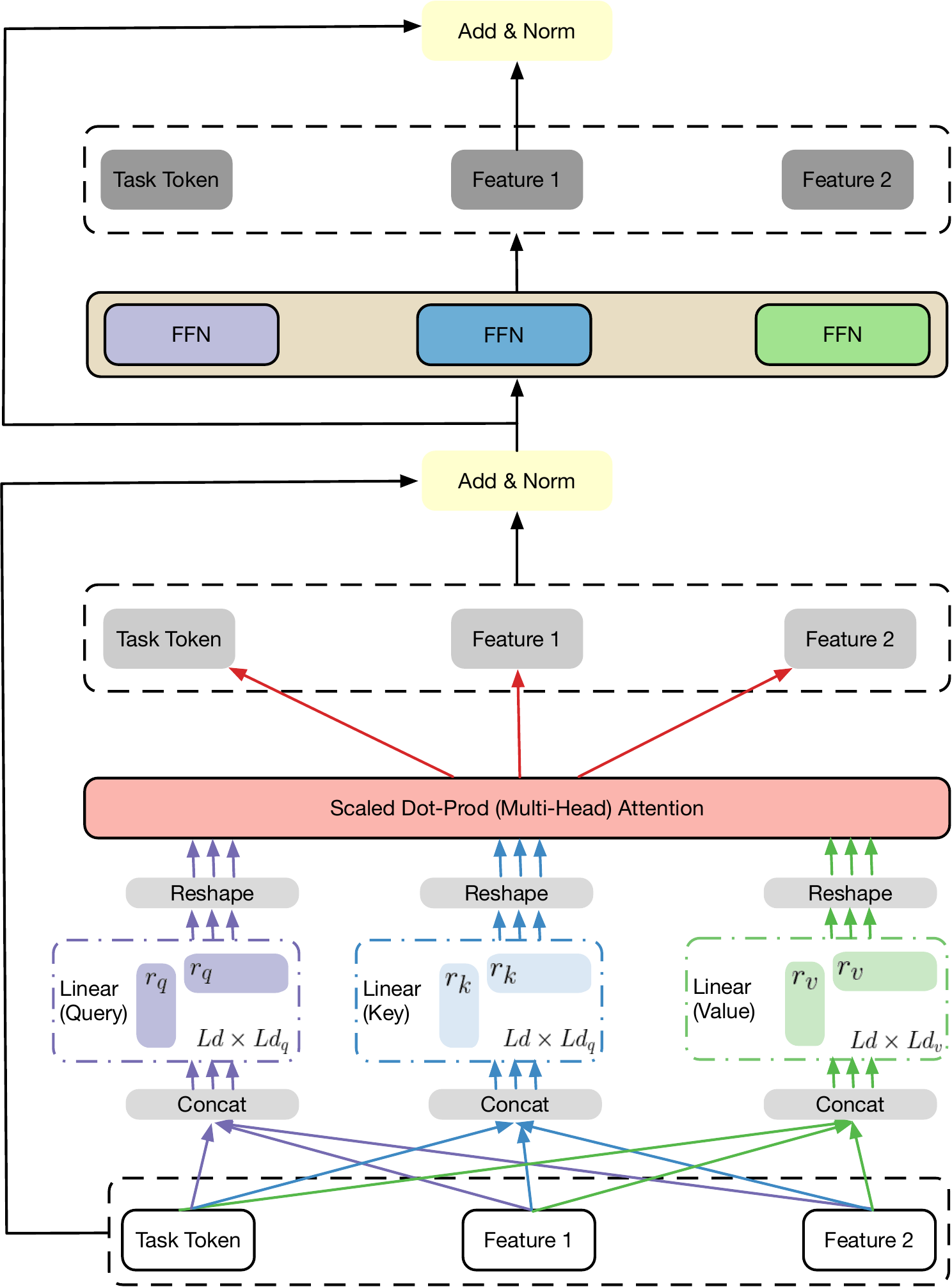}
        \label{fig:ciformer}
    }
    \vspace{-3mm}
    \caption{Illustration on capturing feature interaction with different Transformer-based architectures.
    with one task token and two feature tokens.
    In \ref{fig:fiformer},  apply vanilla Transformer for feature interaction learning. 
    In \ref{fig:hiformer}, we explicitly model the heterogeneous feature interaction through a heterogeneous interaction layer.
    In \ref{fig:ciformer}, we propose the \textsc{Hiformer} model, which further improves the model expressiveness compared with the heterogeneous attention layer in~\ref{fig:hiformer}. $d$ is hidden layer size, $d_q$, and $d_v$ are Query and Value dimensions, $L$ is the total input length,  $r_*$ corresponding to the rank of the corresponding QKV projections (see Eq~\eqref{eq:low-rank-projection}).
    }
    \label{fig:sample_subfigures}
    \vspace{-4mm}
\end{figure*}

Recall that the attention score computation of vanilla Transformer, the parameterization of feature interactions learning are shared across all feature groups in the multi-head self-attention layer.
This parameter sharing design comes naturally in NLP tasks, as the input to a Transformer in NLP tasks is usually a sentence represented by a list of text embeddings.
Text embeddings primarily encode information that is context independent.
We refer this setup as \textit{homogeneous} feature interactions. 
The detailed illustration of the Transformer architecture is shown in Figure~\ref{fig:fiformer}.


By sharp contrast, there is diverse contextual information in recommender systems.
The learned embedding representations are more dynamic and encompass information from multiple perspectives. 
In the food delivery app recommendation example, we have three features: \texttt{app\textunderscore{}id} (\emph{e.g.}, Zomato), \texttt{hour\textunderscore{}of\textunderscore{}day} (\emph{e.g.}, 12pm), and \texttt{user\textunderscore{}country} (\emph{e.g.}, India).
The learned embedding representation  for the app Zomato encodes various  implicit information, including the nature of the app, the countries in which it is popular, the times at which it is frequently used, and the language of its users, etc. 
Not all the encoded information would be useful during certain feature interaction learning.
When learning the feature interaction between \texttt{app\textunderscore{}id}  and \texttt{user\textunderscore{}country}, likely the information of the countries in which the app is popular is much more important than the others. 
Thus, the core of heterogeneous feature interaction learning is to provide contextual awareness for information selection and transformation, to achieve more accurate feature interaction learning.

\subsection{Heterogeneous Attention Layer}
\label{sec::hetero-att}
We first introduce a simple but effective modification to the multi-head self-attention layer in the Transformer model: the heterogeneous  multi-head self-attention layer,  to capture the heterogeneous second-order feature interactions.
For simplicity, we refer to the heterogeneous  multi-head self-attention layer as the heterogeneous attention layer, and the corresponding model as heterogeneous attention model.
In particular, we redesign the multi-head self-attention score computation for feature $i$ and $j$:
\begin{equation}
    \text{Att}(i, j)^h = \frac{\text{exp} \big (\phi^h_{i, j}(\textbf{e}_i, \textbf{e}_j ) \big )}{\sum_{m=1}^L\text{exp} \big (\phi^h_{i, m}( \textbf{e}_i,  \textbf{e}_m ) \big)},
    \label{eq:att-compute}
\end{equation}
where $h$ refers to the $h$-th head among total $H$ heads, $L$ is the embedding list length, $\phi^h_{i, j}(\textbf{e}_i, \textbf{e}_j)$ measures  semantic correlation between embedding $\textbf{e}_i$ and $\textbf{e}_j$ regarding head $h$.
We will then refer the vanilla self-attention layer as homogeneous attention layer.

For every feature pair $(i, j)$, we have a unique  $\phi^h_{i, j}(\cdot, \cdot)$ to measure semantic correlation. 
In other words, $\phi^h_{i, j}(\cdot, \cdot)$ serves the purpose of selecting the most relevant information from $\mathbf{e}_i$ and $\mathbf{e}_j$ for the context of feature pair $(i, j)$. 
We can have arbitrary functions for   $\phi^h_{i, j}(\cdot, \cdot): [\mathbb{R}^d, \mathbb{R}^d] \to \mathbb{R}$, such as some explicit non-additive functions,  a neural network with nonlinear transformation. 
Due to its simplicity, we opt for the dot-product function. 
Accordingly, 

\begin{equation}
\phi^h_{i, j}(\textbf{e}_i, \textbf{e}_j) = \frac{\textbf{e}_i \textbf{Q}^h_i   (\textbf{e}_j \textbf{K}^h_j)^T}{\sqrt{d_k}}, 
\label{eq:att-semantics}
\end{equation}
where $\textbf{Q}_i \in \mathbb{R}^{d \times d_k}$, $\textbf{K}_j \in \mathbb{R}^{d \times d_k}$ are the query and key projections for feature $j$ and $j$ respectively, and $\sqrt{d_k}$ is to normalize magnitude of the dot product, which is often set to be $d_k = d / H$.

With the attention weights computed in Eq~\eqref{eq:att-compute}, we can then compute the output of the heterogeneous attention layer as follows:
\begin{equation}
    \textbf{o}_i = \text{concat}\bigg (\big \{ \sum_{j} \text{Att}(i, j)^h \textbf{e}_j \textbf{V}^h_j\big \}_{h=1}^H \bigg ) \textbf{O}_j,
\end{equation}
where $\textbf{V}_j \in \mathbb{R}^{d \times d_v}$, $ \textbf{O}_j \in \mathbb{R}^{H d_v \times d}$ as value and output projections, and $d_v$ is frequently set as $d_v = d / H$.

Similar to the heterogeneous attention layer, we also design a heterogeneous fully connected Feed Forward Net (FFN) per feature. The FFN is implemented through two linear transformation function with Gaussian Error Linear Unit (GELU) activation~\cite{hendrycks2016gaussian, shazeer2020glu}:
\begin{equation}
\text{FFN}_\text{GELU}^i(\textbf{o}_i) = \text{GELU}(\textbf{o}_i \textbf{W}_1^i + \textbf{b}_1^i) \textbf{W}_2^i + \textbf{b}_2^i,
\end{equation}
where $\textbf{W}_1^i \in \mathbb{R}^{d \times d_f}, \textbf{W}_2^i \in \mathbb{R}^{d_f \times d}, \textbf{b}_1^i \in \mathbb{R}^{d_f},  \textbf{b}_2^i \in \mathbb{R}^d$, and $d_f$ is the layer size of the intermediate layer. 
Following existing studies~\cite{vaswani2017attention}, $d_f$ is set as $d_f = 4d$, $d$ is the dimensionality of model.

As we can see the key change of the heterogeneous attention layer, compared with self-attention layer is that we learn individual query, key, and value (QKV) projection matrices (\textit{i.e.}, $\mathbf{Q, K, V}$) per feature. Therefore, we increase the number of parameters in the heterogeneous attention layer, compared with the vanilla Transformer model. 
The number of parameters scales linearly with the length of input embedding list.
However, it is worth pointing out that the total numbers of FLOPs are the same for the heterogeneous attention layer and homogeneous one (the standard multi-head self-attention layer in the Transformer).
This is because we have exactly the same ops compared with homogeneous Transformer layer, but with different parameterizations.

\begin{figure}
\vspace{-3mm}
    \subfigure[Vanilla Transformer Attention Layer.]
    {
        \includegraphics[height=0.23\textwidth]{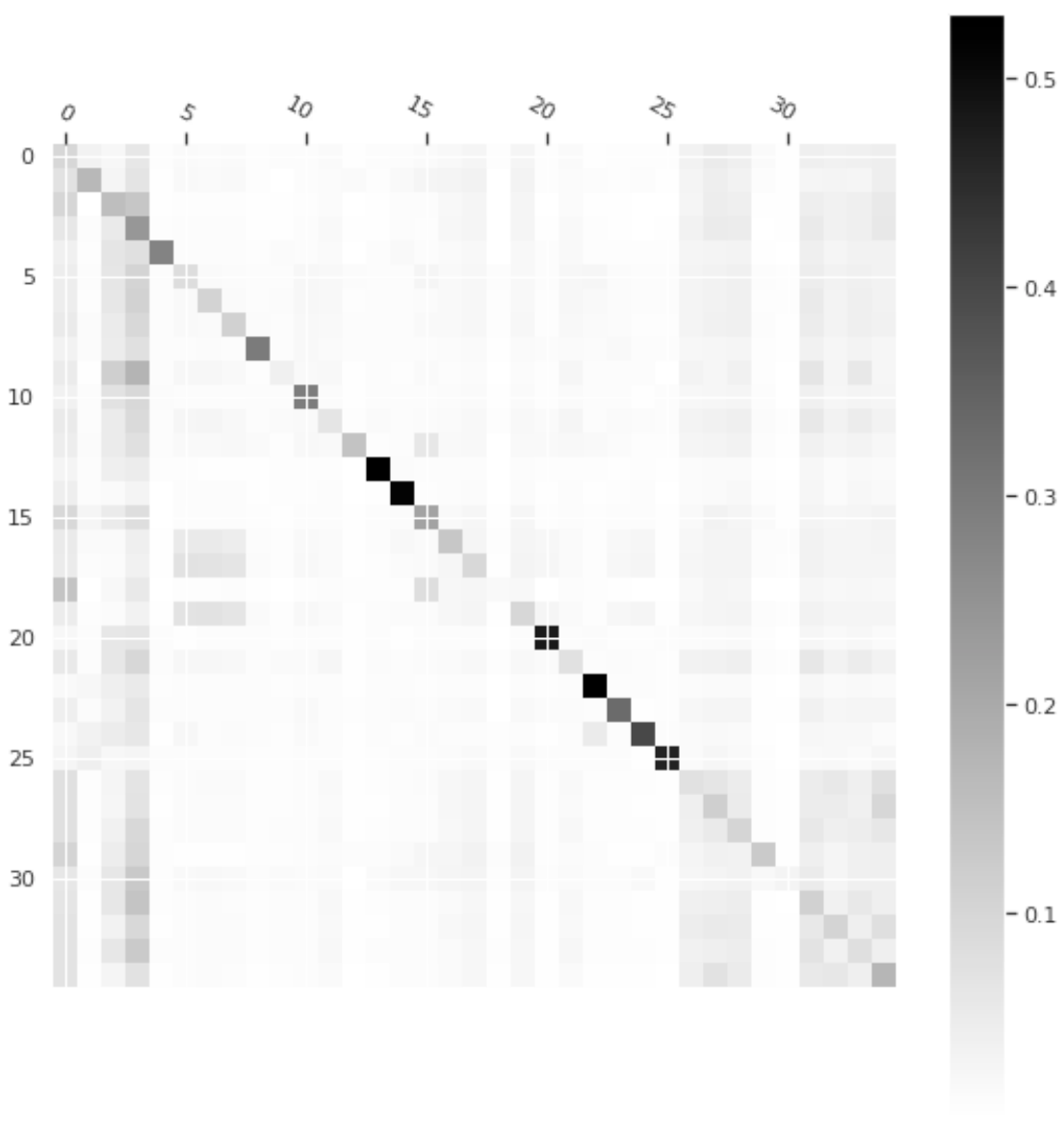}
        \label{fig:att-pattern-fiformer}
    }
          \hspace*{\fill} 
    \subfigure[Heterogeneous Attention Layer.]
    {
        \includegraphics[height=0.23\textwidth]{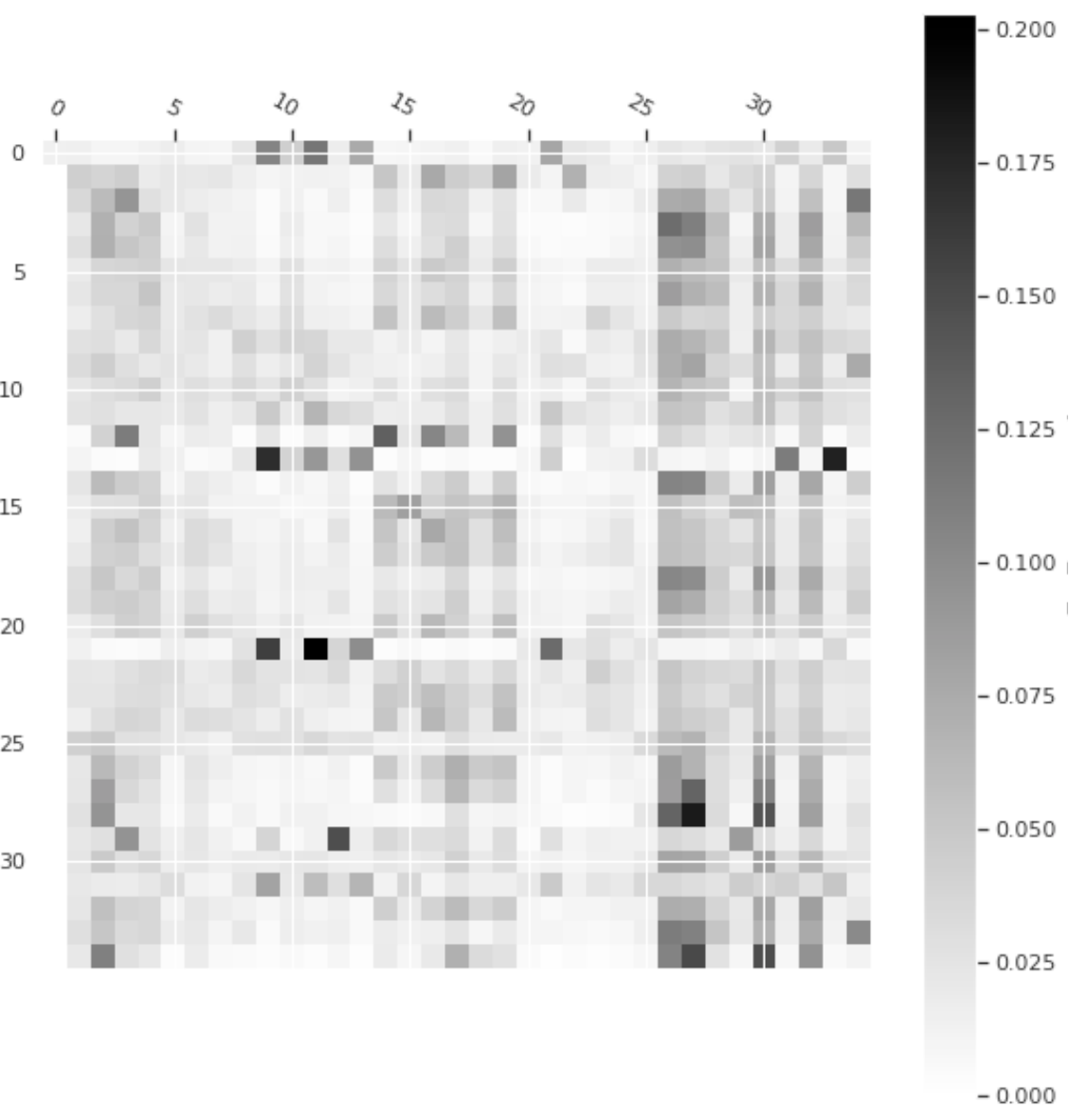}
        \label{fig:att-pattern-hiformer}
    }
    \vspace*{-4mm}
    \caption{Attention patterns with feature interaction layers.}
    \label{fig:att-pattern}
    \vspace{-4mm}

\end{figure}

We visualize the learned attention pattern $\text{Att}(\cdot, \cdot)$  in Figure~\ref{fig:att-pattern}. 
The attention pattern learned with the vanilla Transformer model, in Figure~\ref{fig:att-pattern-fiformer}, is relatively sparse and exhibits a strong diagonal pattern, indicating that the attention layer might not effectively capture the collaborative effects between different features. 
Meanwhile, for the heterogeneous attention layer, there is a dense attention pattern.
Intuitively, the observation could be explained by the fact that the heterogeneous attention layer might be better at capturing the collaborative effects among features thanks to the transformation matrix for feature semantic alignment.

\subsection{\hiformer}
\label{sec::hiformer}

Now we are ready to introduce the \hiformer model, where we further increase the model expressiveness, by introducing composite feature learning in the QKV projections. Take key projection as an example. Instead of learning key feature projection per feature, we redefine the key projection as follows:
\begin{equation}~\label{eq:composite-def}
[\mathbf{\hat{k}}_{1}^h, \ldots \mathbf{\hat{k}}_{L}^h] =  \text{concat}([\mathbf{e}_{1}^h, \ldots \mathbf{e}_{L}^h]) \mathbf{\hat{K}}^{h},
\end{equation}
where $\mathbf{\hat{K}}^{h} \in \mathbb{R}^{Ld \times Ld_k}$.
Compared with key projection matrices in Section~\ref{sec::hetero-att}, $[K_i^h, \ldots, K_L^h] \in \mathbb{R}^{d \times Ld_k}$, we increase model expressiveness by adding more parameters to key projections.
We call the new projection $\mathbf{\hat{K}}^{h}$ as \textsc{Composite} projection. 
Mathematically, instead of learning feature interactions explicitly with feature pairs, we first transform the feature embedding list into composite features as keys and values; then we learn heterogeneous feature interaction between composite features and task embeddings. 
Similarly, we apply cross projection to query and value projections.

Similarly as in Eq~\eqref{eq:att-compute}, the \hiformer model computes the attention scores between the query $\mathbf{q}_i$ and and key $\mathbf{k}_i$ 
as follows:
\begin{equation*}
    \text{Att}_\textsc{Composite}(i, j)^h =  \frac{\text{exp} \big ( \mathbf{\hat{q}}_i^h ( \mathbf{\hat{k}}_i^h )^T / \sqrt{d_k} \big ) }{\sum_{l}^L \text{exp} \big ( \mathbf{\hat{q}}_i^h ( \mathbf{\hat{k}}_l^h )^T / \sqrt{d_k} \big) },
    \label{eq:cross-att-compute}
\end{equation*}



With the attention score, we can then compute the output of attention layer in \textsc{Hiformer} model as:
\begin{equation}
        \mathbf{\hat{o}}_i = \text{concat}\bigg (\big \{ \sum_{m} \text{Att}^h_\textsc{Composite}(i, j) \mathbf{\hat{v}}_i^h \big \}_{h=1}^H \bigg ) \mathbf{{O}}_j.
\end{equation}

\subsection{Better Efficiency for \hiformer}
Compared with the Transformer model and the heterogeneous attention layer, the \hiformer model comes with more model expressiveness, but also computation cost during inference. 
Due to the introduced \textsc{Composite} attention layer, the existing efficient Transformer architectures~\cite{tay2022efficient} cannot be directly applied.

The length of the feature embedding list is $L = |\mathcal{C}| +  n^\mathcal{D} + t$.
With the model dimension as $d$, we have inference cost breakdown~\footnote{This computation is based on the setting where $d_k = d / H, d_v = d / H$, and $d_f = 4d$.}  for \textsc{Hiformer}: 
QKV projection: $3L^2d^2$, attention score computation: $2L^2d$, output projection: $Ld^2$, FFN network: $8Ld^2$. As a result, total computation is $\mathcal{O}(L^2d^2 + L^2d + Ld^2)$. 
To deploy \hiformer for real-time inference, we reduce the inference cost of \hiformer through  low-rank approximation and pruning.

\subsubsection{Low-rank approximation}
\label{subsec:low-rank-approximation}
As we can see, the outstanding term for \hiformer inference cost is QKV projections with computation complexity of $\mathcal{O}(L^2d^2)$. 

\begin{figure}[H]
\vspace{-3mm}
        \includegraphics[width=0.25\textwidth]{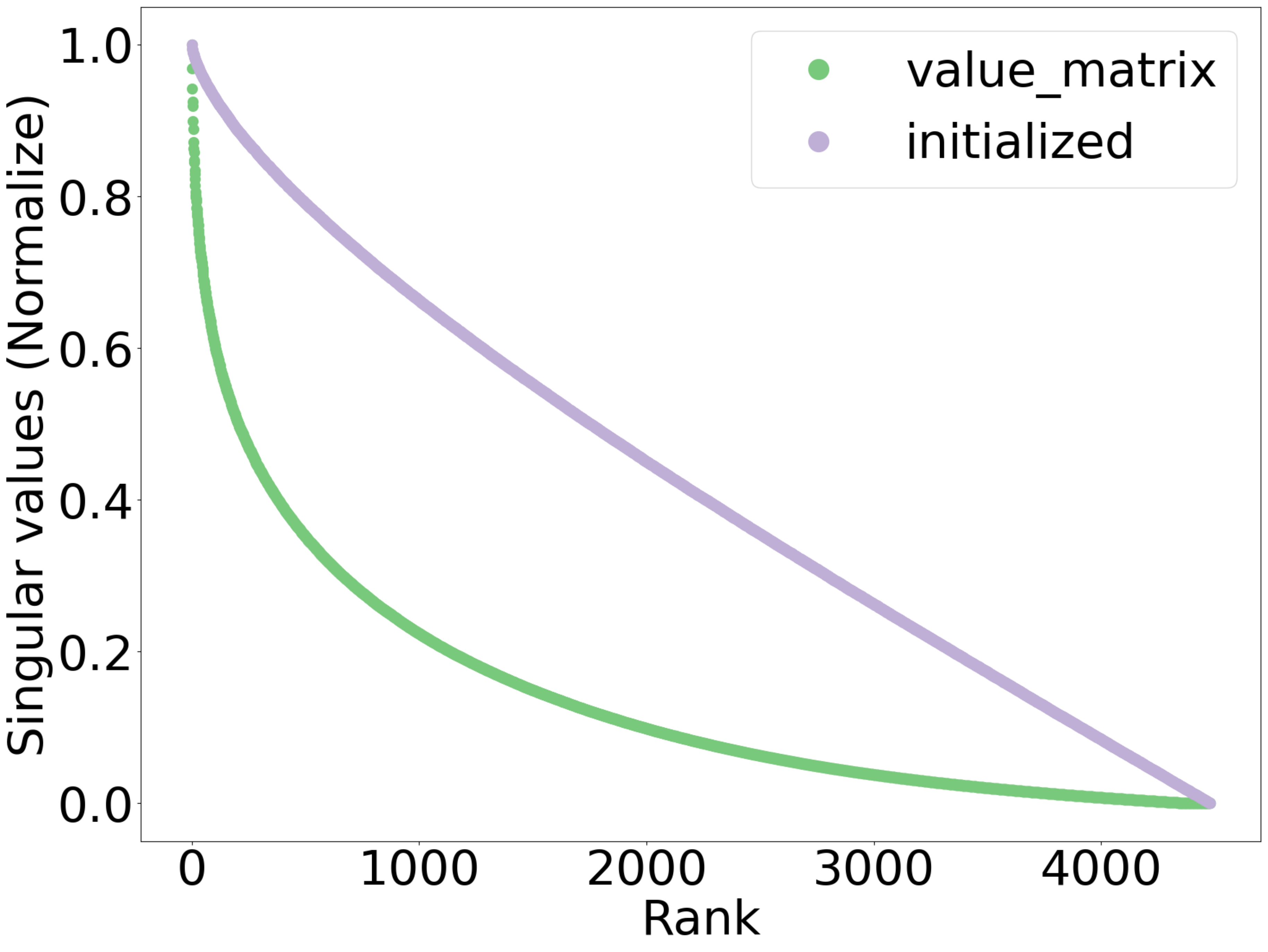}
        \label{fig:cross-k-v-projection}
    \vspace*{-2mm}
    \caption{Singular values for $\mathbf{\hat{V}^h}$ in \textsc{Composite} projection.}
    \label{fig:cross-k-v-low-rank}
    \vspace{-3mm}

\end{figure}

With \textsc{Composite} key projection defined in ~\eqref{eq:composite-def}, we can approximate $\mathbf{\hat{K}}^h$ with low-rank approximation, i.e.,
\begin{equation}
    \mathbf{\hat{K}^h} = \textbf{L}_k^h (\textbf{R}_k^h)^T, ~\label{eq:low-rank-projection}
\end{equation}
where $\textbf{L}_v^h \in \mathbb{R}^{Ld \times r_v}$, $\textbf{R}_v^h \in \mathbb{R}^{Ld_k \times r_v}$, and  $r_v$ is the rank of the low-rank approximation for $\mathbf{\hat{K}}^h$.  
Low-rank approximation reduces the computation cost for value projection to $\mathcal{O}(Lr_v(d + d_k))$. Similarly, it can be applied to query and value \textsc{composite} projection with rank of $r_k$ and $r_v$.
If $r_k < Ld_k / 2$ and $r_v < Ld_v / 2$, the cost will be reduced. The low-rank structure is also observed in data analysis.
Taking $\mathbf{\hat{V}}^h$ as an example, we plot the singular values in Figure~\ref{fig:cross-k-v-low-rank}, where we can see that the  $\mathbf{\hat{V}}^h$  matrix is indeed low-rank.

After applying the low-rank approximation, the \hiformer model complexity is computed as follows:
 the query and key projection: $\mathcal{O}(L r_k(d + d_k))$, value projection  $\mathcal{O}(L r_v (d + d_v))$, attention layer $\mathcal{O}(L^2d)$, output projection $\mathcal{O}(Ld^2)$, and FFN layer $\mathcal{O}(Ld^2)$.
 With $d_k = d / H$, $d_v = d / H$, and $r_k < d_k, r_v < d_v$, we have the computation complexity as $\mathcal{O}(L^2 d + Ld^2)$, which scales quadratically with the length of the input embedding list $L$.

\subsubsection{Model Pruning.}
Recall that in the Output Layer (see Figure~\ref{fig:framework}), we are only using the encoded task embeddings for final objective training and predictions. 
Therefore, we can save some computation in the \textit{last layer} of the \hiformer model by pruning the computation of encoding feature embeddings.
In particular, the pruning translates to only using task embeddings as query, and using both task embedding and feature embeddings as key and value in the \hiformer attention layer. 
Therefore, we have   computation complexity of the last \hiformer layer:
query projection: $\mathcal{O}(t r_k(d + d_k))$, key projection: $\mathcal{O}(L r_k(d + d_k))$, value projection: $\mathcal{O}(L r_v (d + d_v))$, attention layer computation: $\mathcal{O}(Ltd)$, output projection: $\mathcal{O}(td^2)$, FFN layer: $\mathcal{O}(td^2)$. 
Suppose the number of task $t$ ($\ll L$) is considered to be a constant.
We have the computation complexity of the last layer of \textsc{Hiformer}  as $\mathcal{O}(L(r_k+r_v)d + Ltd + td^2)$.

 Pruning enables us to reduce the computation complexity of the last \hiformer layer from scaling quadratically to scaling linearly with $L$.
 This pruning technique can be applied to all Transformer-based architectures.
 Some of the recent work , such as Perceiver~\cite{jaegle2021perceiver}, has also introduced a similar idea to reduce the serving cost of Transformer models from scaling quadratically to linearly.

\section{EXPERIMENT}
\label{sec::experiment}
We undertake a thorough empirical examination of the proposed model architectures with the web-scale App ranking model at Google Play.
We conduct both offline and online experiments, with which we aim at answering the following research questions:
\begin{itemize}
 \item[\textbf{Q1.}] Is the heterogeneous attention layer  able to capture the heterogeneous feature interactions to give better recommendations, compared with vanilla Transformer models? 
 \item[\textbf{Q2.}] With more model expressiveness in the \hiformer~~model, could it further improve model quality?
 \item[\textbf{Q3.}] How is the serving efficiency of  the  \textsc{Hiformer} model compared with SOTA models? 
 \item[\textbf{Q4.}] How do the hyper-parameter settings impact the model performance and serving latency and how to make the performance and efficiency trade-offs?
\end{itemize}

\subsection{Offline Evaluation Setup}

\subsubsection{Dataset}
For offline evaluation, we use the logged data from the ranking model.
Users' engagements are training labels, either 0 or 1 indicating if the engagement happens between user and the app and the training loss is LogLoss (see Eq~\eqref{eq:logloss}).
We train the model with 35 days of rolling-window data, and evaluate on the 36th day data. 
The model is retrained from scratch everyday with most recent data.
There are 31 categorical features and 30 dense scalar features to describe the apps, such as app ID, app title, user language, etc.
The vocabulary size of the categorical features varies from hundreds to millions. 

\subsubsection{Evaluation Metrics}
We will evaluate the model performance using AUC (Area Under the Receiver Operating Characteristic Curve).
The AUC metric can be interpreted as the probability of the model assigning a randomly chosen positive example a higher ranking than a randomly chosen  negative example.
In the web-scale application dataset, due to the large size of the evaluation dataset, we notice that the improvement of AUC at 0.001 level achieves statistical significance. Similar observations have been made in other web-scale applications~\cite{cheng2016wide, guo2017deepfm, joglekar2020neural}. Additionally, we report the normalized LogLoss with Transformer one layer as the baseline.

We train the model on TPU~\cite{jouppi2017datacenter, jouppi2018motivation}, and we report the training Queries per second (QPS) to measure the training speed.
We  estimate the latency through offline simulation, where we perform inference of the models on 20 batches of examples with batch size being 1024. 
We then normalize the latency of all models with respect to the baseline model, which is a one-layer Transformer model with pruning.
Due to the large number of online serving requests, the serving latency is much more critical than training QPS.



\subsubsection{Model Architectures}

Note that our work is on feature interaction learning with Transformers, which is drastically different from user history sequence modeling with Transformers~\cite{kang2018self,sun2019bert4rec}. Therefore, we focus on comparing our proposed models with the following SOTA feature interaction models:
\begin{itemize}[leftmargin=*]
    \item \textbf{AutoInt}~\cite{song2019autoint} is proposed to learn feature interaction through the multi-head self-attention layer, and then learn the final predictions through a Multi-layer Perceptrons (MLP) layers.
    \item \textbf{DLRM}~\cite{naumov2019deep} is to learn feature interaction through factorization machine, and then learn implicit feature interactions through a MLP layer, for the final predictions.
    \item \textbf{DCN}~\cite{wang2017deep} (\textit{i.e.,} DCN-v2) is  one of the SOTA models on feature interaction learning in web-scale application, which has a cross network that explicitly creates bounded-degree feature crosses where the feature interaction order increases with layer depth.
    \item \textbf{Transformer}~\cite{vaswani2017attention} with multi-head self-attention layer for feature interaction learning. The main difference between Transformer and AutoInt is: 1. Transformer use encoded task embeddings for predictions; while AutoInt uses all encoded feature and task embeddings; 2. There is FFN layer in Transformer architecture, but not in the AutoInt layer.
    \item \textbf{Transformer + PE} is to add  \textbf{P}osition \textbf{E}ncoding to the Transformer architecture, where we learn an additional per-feature embedding as feature encoding.
    \item \textbf{HeteroAtt} is  our proposed Heterogeneous Attention layer to capture the heterogeneous feature interactions.
    \item \textbf{\textsc{Hiformer}} is our proposed model with low-rank approximation and model pruning.
\end{itemize}




\subsubsection{Implementation Details}
The model training and serving is built on Tensorflow 2~\cite{singh2020introduction} and Model Garden~\cite{yu2020tensorflow}.
To ensure a fair comparison, we keep all the model components, except for the feature interaction layers, to be the same. 
In the embedding layer, we set the embedding dimensions to be 128 for all models. 
Note that for Transformer architectures, HeteroAtt, and \hiformer, we only use the encoded CLS token (i.e., learned task token) for the final task predictions. 

In our experiments on Transformers, we set model dimensionality $d = 128$, number of heads in the attention layer $H = 4$, and the layer size of FFN to be $d_f = 512$. 
We set $d_k = 16$ and $d_v = 64$, instead of setting $d_k = d / H, d_v = d / H$ as in the Transform model and the other Transformer-based models. 
For the other methods, we perform a hyper-parameters tuning to get the best performance.
For online experiments, we tune the hyperparameters based on one day's (say Feb 1st) data, and fix the settings for the following days (i.e., dates after Feb 2nd).

\subsection{Homogeneous vs Heterogeneous Feature Interaction (Q1)}
First of all, we observe that Transformer+PE model performs similarly to the vanilla Transformer model. This is consistent with our hypothesis. Though coming with more parameters, the feature encoding only learns a bias embedding term per feature, which is not sufficient to learn the feature transformation for heterogeneous feature interaction learning.
Secondly, we observe that Transformer model with heterogeneous attention layer  (HeteroAtt) performs significantly better than the vanilla Transformer model. 
This result validates our argument that HeteroAtt can effectively capture the complex feature interactions through the transformation matrices $\textbf{M}_{i, j}$.
Moreover, we observe that the parameter \# of the two-layer HeteroAtt model is much larger than the two-layer Transformer  model, as the HeteroAtt model has more model expressiveness.
Therefore,  we can answer affirmatively to Q1, that the heterogeneous feature interaction with better feature context awareness is crucial for providing personalized recommendations.

\begin{table*}[t]
\begin{tabular}{c c c c c c c } 
 \hline
 Model  & Layer \# & Parameter \# &  AUC ($\uparrow$)  &  LogLoss (\%, $\downarrow$) & Train QPS ($\uparrow$)  & Serving Latency ($\downarrow$) \\ 
 \hline
  AutoInt  & 1  & 12.39M & 0.7813 & -0.37096 & 5.45e6  & 2.28 \\ 
 \hline
  DLRM & - & 5.95M  & 0.7819  & -0.47695 & 5.14e6   & 0.95 \\ 
 \hline
  DCN & 1 & 13.73M & 0.7857 & -0.79491  & 5.76e6 & 1.46 \\ 
 \hline
 \multirow{3}{7.5em}{{Transformer}}  &  1 & 0.74M  & 0.7795 & 0 & 5.75e6  &  1.00 \\
  &  2 & 0.84M  & 0.7811 & -0.31797 &  4.12e6 &  3.03  \\
 &  3  & 0.97M  & 0.7838 & 	-0.45045 &  3.13e6 &  5.05 \\
 \hline
 \multirow{1}{7.5em}{{Transformer+PE}} &  3  & 1.08M  & 0.7833  & 	-0.39746 &  3.12e6 &  5.06 \\
 \hline
 \multirow{2}{7.5em}{HeteroAtt (ours)}  &  1& 2.36M  & 0.7796  & -0.05299 &  5.71e6  &  1.01 \\
 &  2  & 10.50M  & 0.7856   & -0.82141 &   4.10e6  &  3.11 \\
  \hline
 \multirow{1}{7.5em}{\textsc{Hiformer}  (ours)} &  1 &  16.68M  & 0.7875  & -0.87440 &  5.69e6  & 1.52  \\
 \hline

\end{tabular}
\caption{Offline model performance comparison regarding number of parameters, AUC, and normalized efficiency. The number of parameters excludes the embedding layer. $\uparrow$ ($\downarrow$) means the higher (lower) the better.}
\vspace{-7mm}
\label{tbl:all-res}
\end{table*}

\subsection{Model Performance Comparison (Q2)}
The offline comparison of the baseline models and our proposed models are summarized in Table~\ref{tbl:all-res}. 
We report the best performance of each model, with the corresponding number of layers. 

We take a closer look at the AutoInt, DLRM, and vanilla Transformer model. 
In the feature interaction learning of these models, there is no feature awareness and semantic alignment. 
Thus, their performance are relatively weaker compared with the other models. 
In DCN model, the Cross Net implicitly generates all the pairwise crosses and then projects it to a lower-dimensional space. The parameterization of all pairwise crosses are different, which is similar to providing feature awareness. 
Therefore, DCN model performance is comparable with our HeteroAtt models.

Compared with the heterogeneous attention layer (i.e., HeteroAtt),  \hiformer model provides more model expressiveness in the QKV projections. It achieves the best model performance with just one layer, outperforming the HeteroAtt model and the other SOTA models.

\subsection{Serving Efficiency (Q3)}
We compare the serving efficiency of all the models.
First of all, we compare the Transformer two layer model with the one layer model. As we can see, the latency of the two layer model is more than 2x of the one layer model. This is due to the fact that we only apply pruning to the second layer in the two layer model, and we have already applied pruning to the one layer model.
Similarly, we can draw the same conclusion for the HeteroAtt layer as the two models have the same operators, thus latency.
However, such pruning technique cannot be applied to AutoInt model. Therefore, the one-layer AutoInt model is much more expensive than the vanilla Transformer model and the HeteroAtt model. 

Secondly, we compare the \hiformer model with the one-layer HeteroAtt model. 
Both the one-layer \hiformer model and the one-layer HeteroAtt model enjoy the inference optimization from pruning. However, because of the more expressive QKV projection, we see that the \hiformer model is 50.05\% more expensive. 

Nevertheless, we want to point out that without low-rank approximation, the \hiformer model would potentially be much more expensive. 
In our experiment, we set rank of query and key as $r_k = 128$, and value $r_v = 1024$.
We compare the \hiformer with and without low-rank approximation in Table~\ref{tab:low-rank-latency}.
The low-rank approximation gives 62.7\% inference latency saving.
Additionally, there is no significant model quality loss, which is expected because of the observed low-rank structures of \hiformer query, key, and value projection matrices (\textit{e.g.}, $\mathbf{\mathcal{V}}$, shown in Figure~\ref{fig:cross-k-v-low-rank}).

\begin{table}[H]
    \centering
    \begin{tabular}{c  c c c}
    \hline
        \hiformer Model & Parameter \# &  AUC & Latency  \\
    \hline
        low-rank approx. &  16.68M & 0.7875 &  1.52 \\
    \hline
        w/o low-rank approx. & 59.95M &  0.7882 & 3.35 \\
    \hline
    \end{tabular}
    \caption{Low-rank approximation for \hiformer.}
    \label{tab:low-rank-latency}
\end{table}


\subsection{Parameter Sensitivity (Q4)}
Since one of the challenges is to reduce the serving latency, we are also interested in the parameter sensitivity of the model as in RQ4. More formally, we aim to answer the following question:

\begin{figure}[b]
    \subfigure[HeteroAtt: $Hd_k$.]
    {
        \includegraphics[scale=0.07]{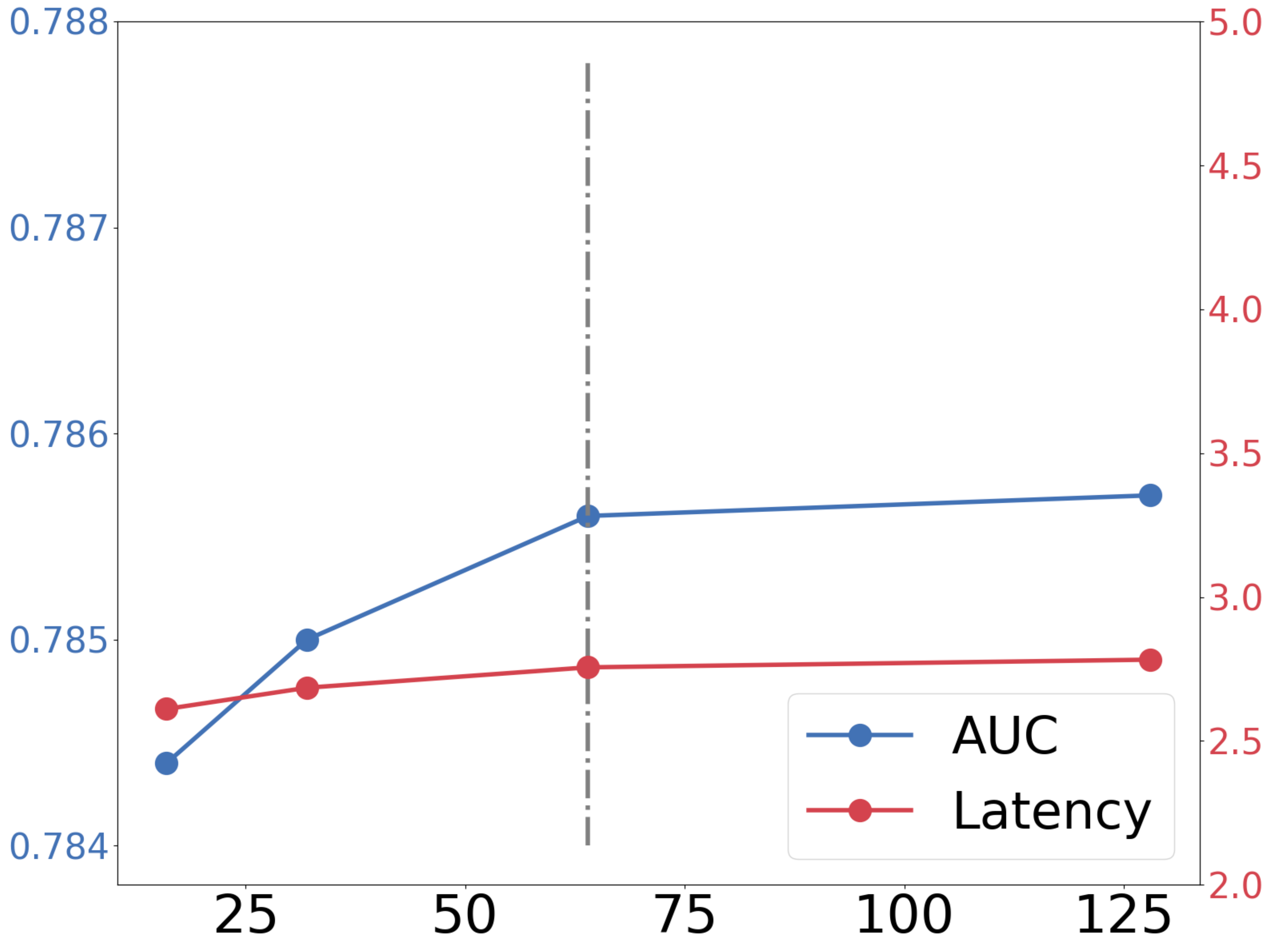}
        \label{fig:hetero-dk}
    }
    \subfigure[HeteroAtt: $Hd_v$.]
    {
        \includegraphics[scale=0.07]{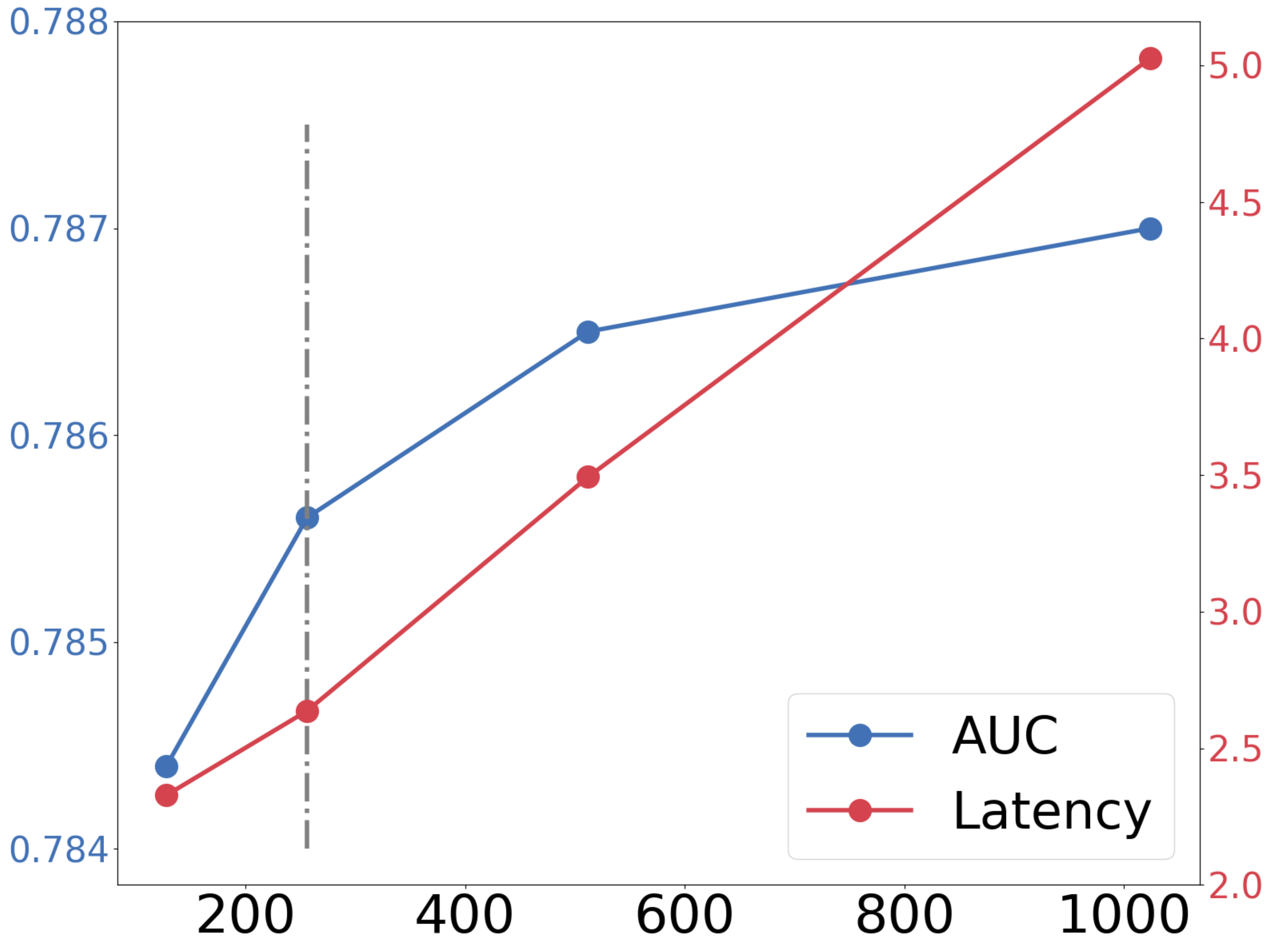}
        \label{fig:hetero-dv}
    }
        \subfigure[\hiformer: $Hd_k$.]
    {
        \includegraphics[scale=0.07]{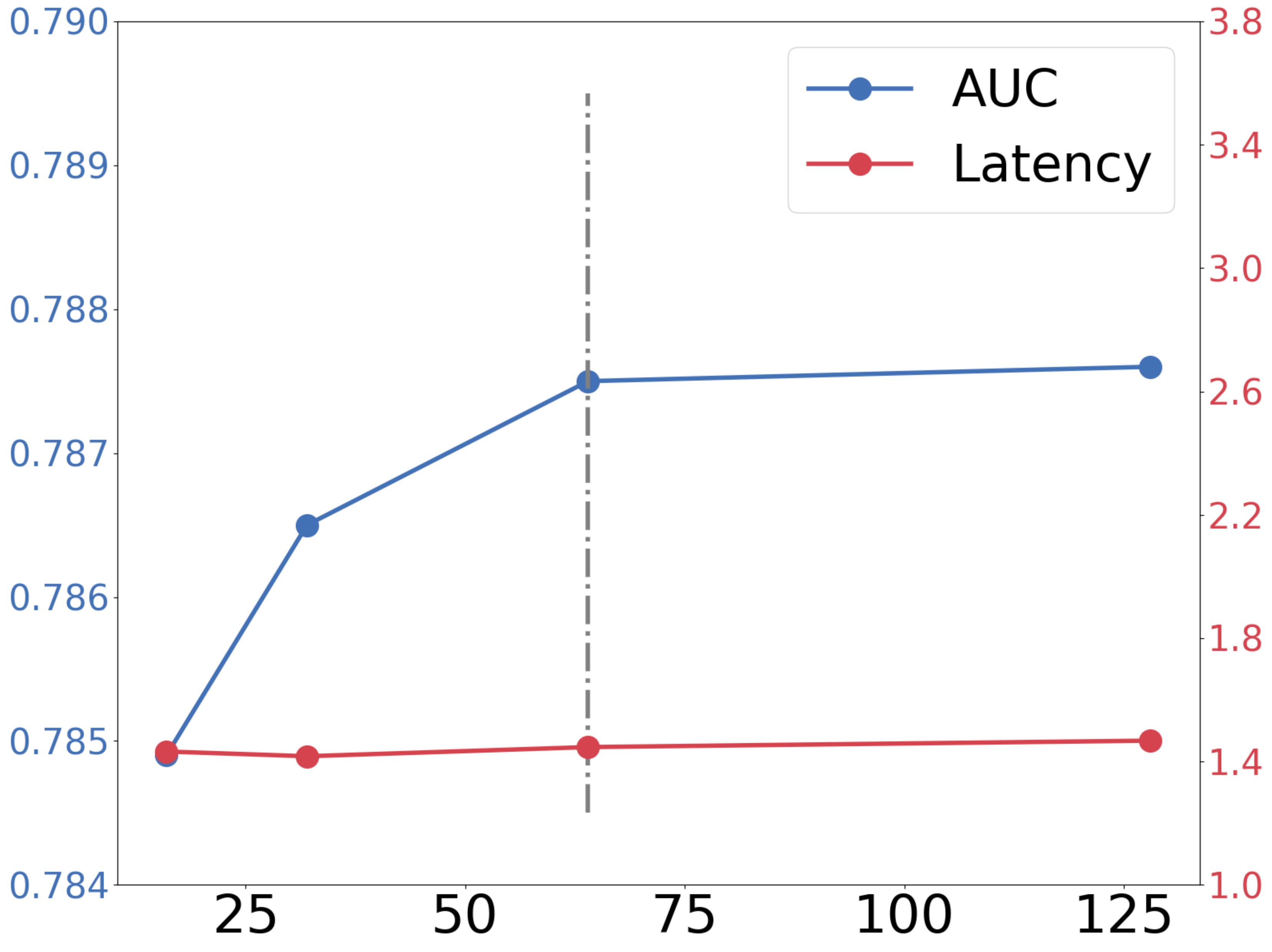}
        \label{fig:hiformer-dk}
    }
    \subfigure[\hiformer: $Hd_v$.]
    {
        \includegraphics[scale=0.07]{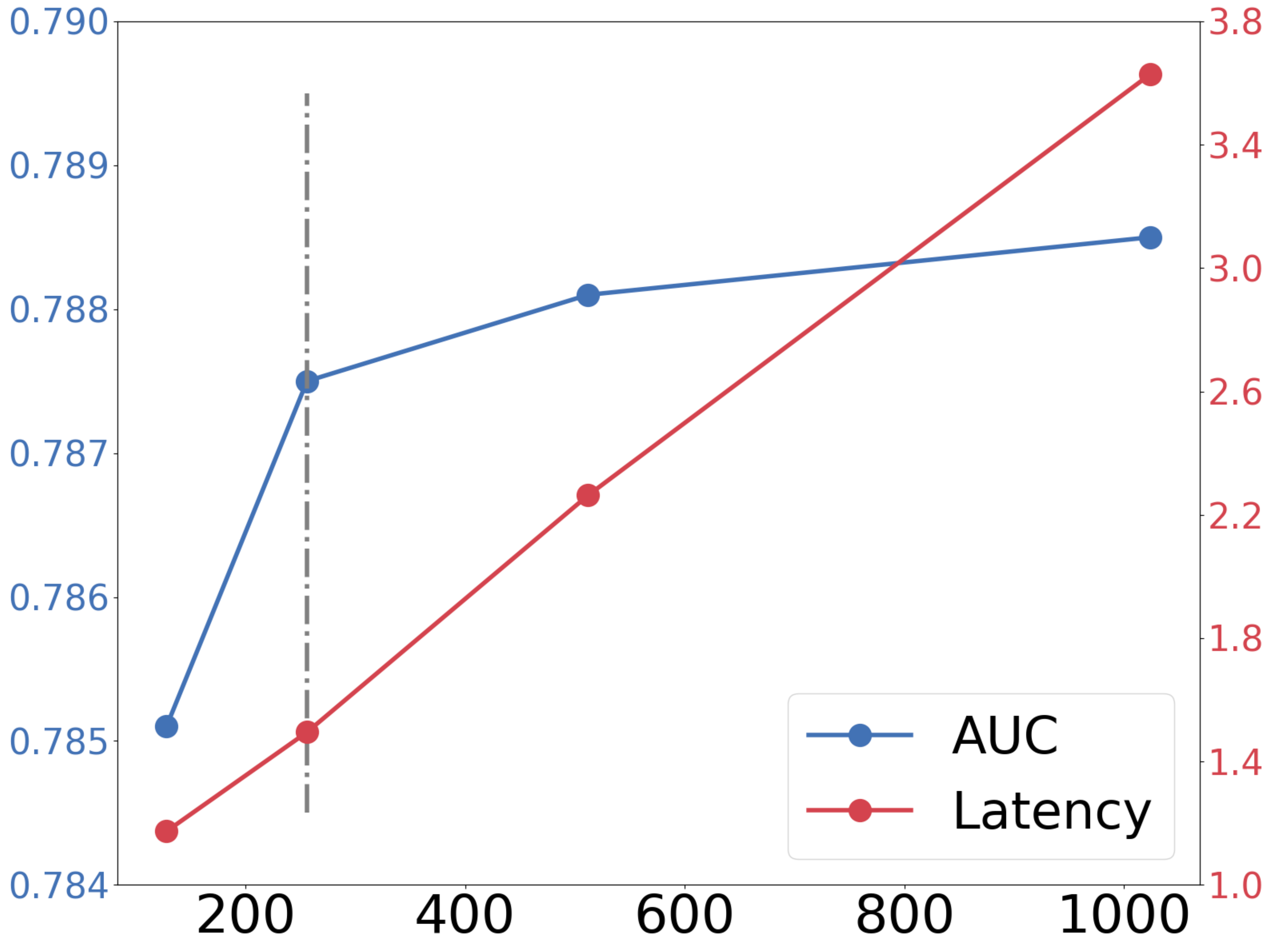}
        \label{fig:hiformer-dv}
    }  
    \vspace{-5mm}
    \caption{Parameter Sensitivity. With increased $d_k$ and $d_v$, the model quality gain saturates; however, the model inference cost increases drastically. The black dotted lines mark the selected $d_k$, $d_v$ values, which gives the best quality and inference cost trade-off.}
    \label{fig:sensitivity}
\end{figure}

As we previously discussed in Section~\ref{sec::experiment}, instead of directly using $d_k = d / H$ and $d_v = d / H$, we find that the choice of $d_k$ and $d_v$ presents certain trade-offs between model performance and serving latency for the HeteroAtt and \hiformer model. 

Firstly, for $Hd_k$ in HeteroAtt in Figure~\ref{fig:hetero-dk}, after reducing the $Hd_k$ from $d = 128$ to 64, there is no model quality loss. However, we can harvest a free latency improvement (3\%). If we further decreasing $Hd_k$, there is considerable model quality loss. Therefore, we opt for $Hd_k = 64$, as highlighted by the grey line.
By comparison, increasing $Hd_k$ for HeteroAtt model gives model quality improvement, but also significant latency increasing (in Figure~\ref{fig:hetero-dv}). For quality and latency trade-off, we set $Hd_v = 256$. 

Similarly, we observe in Figure~\ref{fig:hiformer-dk}, for \hiformer model, decreasing $Hd_k$ to 64 gives no quality loss, but unlike HeteroAtt model, there is almost no latency gain. 
This is due to the low-rank approximation in the QK projection, especially we set the rank $r_k = 256$, which is relatively small, compared with $r_v = 1024$.
Accordingly, the query projection dominates the computation cost.
With increased $Hd_v$, the latency increases significantly, shown in Figure~\ref{fig:hiformer-dv}.

This observation on the model quality and latency trade-off of $Hd_v$ provides uniqueness of the HeteroAtt and \hiformer model.
It can serves as a instrument to tune the model capacity at the cost of latency. 
For example, when the number of requests per query is relatively small, we can achieve better model performance by increasing $d_v$.
Such an instrument might not not available for the other SOTA models, such as DCN~\cite{wang2021dcn}.


\subsection{Online A/B Testing}
In addition to the extensive offline experiments, we also have conducted an online A/B testing experiment with the proposed model framework. 
For each of the control and treatment groups, there is 1\% randomly selected users, who receive the item recommendations based on different ranking algorithms. 
To understand the improvement we gain through heterogeneous feature interaction learning, the baseline model is leveraging a one-layer Transformer model to capture feature interactions. 
We collect the users' engagement metrics for 10 days, as shown in Table~\ref{tab:online}. 
The one layer HeteroAtt model significantly outperforms the one layer Transformer model, which is consistent with our offline analysis. 
Additionally, two layer HeteroAtt model further improves over the one-layer one.
The \textsc{Hiformer} performs the best among all models, including the DCN model.
Such an observation indicates that Transformer-based architectures can outperform SOTA feature interaction models.
We have successfully deployed the Hiformer model to production.

\begin{table}[h]
    \centering
    \begin{tabular}{c  c c}
        Model & Layer Num & Engagement Metrics \\ \hline
         Transformer  & 1 &  +0.00\%   \\ \hline
         HeteroAtt (ours)  & 1  &  \textbf{$\mathbf{+1.27\%^*}$}    \\ \hline
        HeteroAtt (ours)   & 2 &  \textbf{$\mathbf{+2.33\%^*}$}   \\ \hline
         \textsc{Hiformer}  (ours) & 1 &  \textbf{$\mathbf{+2.66\%^*}$}   \\ \hline
          \textsc{DCN} & 1 & \textbf{$\mathbf{+2.20\%^*}$}  \\ \hline
    \end{tabular}

    \caption{The online engagement and serving latency metrics of different models. $^*$ indicates statistically significant.}
    \label{tab:online}
    \vspace{-10mm}
\end{table}

\section{RELATED WORK}
\label{sec::related}

Before the era of deep learning, people often manually crafted cross features and added them to logistic regression models to improve their performance. Some also leveraged decision tree based models~\cite{he2014practical} to learn feature crosses. Later on, the development of embedding techniques have led to the design of Factorization Machines (FMs)~\cite{rendle2010factorization}, which was proposed to model second-order feature interactions with the inner product of two latent vectors. 

Recently, DNNs have become the backbone of many models in industry. To improve the efficiency of feature cross learning, many works explicitly model feature interactions by designing a function $g(x_i, x_j)$ while leveraging the implicit feature crosses learned by DNNs.
Wide \& deep model~\cite{cheng2016wide} combines a DNN model with a wide component that consists of crosses of raw features. Since then, many work has been proposed to automate the manual feature cross work of wide \& deep by introducing FM-like operations (inner products) or Hadamard products.
DeepFM~\cite{guo2017deepfm} and DLRM~\cite{naumov2019deep} adopt FM in the model, Neural FM~\cite{he2017neural} generalizes FM by incorporating the Hadamard product, PNN~\cite{qu2016product} also uses inner products and outer products to capture pairwise feature interactions.
These methods however can only model up to second order feature crosses.
There is also work that can model higher-order feature interactions. Deep Cross Network (DCN)~\cite{wang2017deep, wang2021dcn} designs a cross network to explicitly learn high-order feature interactions where its interaction order increases with layer depth.
DCN-V2~\cite{wang2021dcn} makes DCN more expressive and practical in large-scale industrial settings.
xDeepFM~\cite{lian2018xdeepfm} also improves the expressiveness of cross network in DCN and relies on the Hadamard product to capture high-order feature crosses.

Another line of work leverages the attention mechanism in Transformer models to capture feature interactions.
AutoInt~\cite{song2019autoint} was proposed to capture the feature interaction through the multi-head self-attention layer. 
Nevertheless, as the Transformer was designed to model relationships that are context independent, such as text tokens, directly applying the multi-head self-attention layer for feature interaction has very limited model expressiveness. 
Similarly,~\cite{li2020interpretable} is based on the attention mechanism with cross-product transformation to capture hierarchical attention, without providing feature awareness and semantic alignment. 


The Heterogeneous Graph Transformer~\cite{hu2020heterogeneous} work is related to ours, where  node types are considered in the heterogeneous mutual attention layer. 
Though we share some similarities in the motivations, our work is substantially different because  \hiformer model is proposed to capture feature interactions  for web-scale recommender system, and the design of the \hiformer attention layer has more expressiveness.
The field-aware factorization machine~\cite{juan2016field} work is also relevant. In~\cite{juan2016field}, the feature awareness of feature interaction learning is implemented through individual embedding lookup tables, while our proposed methods are through the information transformation in the heterogeneous attention layer.
PNN~\cite{qu2018product} leverages kernel Factorization Machine to consider feature dynamics in feature interaction learning, which shares similarity with the heterogeneous attention layer. 
Nevertheless, heterogeneous attention layer is attention based while PNN is derived from factorization machine.
Additionally, we leverage Query and Key projection to learn heterogeneous feature interactions while PNN is based on kernel matrices.
Moreover, we further increase model expressiveness in \hiformer with cost reduction.


\section{CONCLUSION AND FUTURE WORK}
\label{sec::conclusion}

In this work, we propose a heterogeneous attention layer, which is a simple but effective modification of the self-attention layer to provide feature awareness for feature interaction learning.
We then further improve the model expressiveness, and propose \hiformer to improve feature interaction learning.
We also improved the serving efficiency of \hiformer such that it could satisfy the serving latency requirement for web-scale applications. 
The key change in our method is to identify the different relationships among different features in the attention mechanism, such that we can capture semantic relevance given the dynamic information in learned feature embeddings.
Both offline and online experiments on the world leading digital distribution service platform demonstrate the effectiveness and efficiency of our method. 
These experiment results indicate that a Transformer based model for feature interaction learning can perform better than SOTA methods in web-scale applications. 

Our work serves as a bridge to bring Transformer-based architectures to applications in web-scale recommender system.
For future work, we are interested in how the recent advances in Transformer architectures~\cite{han2022survey} in other domains domains, such as NLP and CVR,  can be translated to recommender systems. 
With the new feature interaction component introduced in this paper, it can also be leveraged in neural architecture search~\cite{elsken2019neural} to further improve recommender systems. 


\newpage 

\bibliographystyle{acm}
\bibliography{sigproc}

\newpage

\appendix

\end{document}